\documentclass[a4paper, 12pt]{article} 
\usepackage[margin=20mm, top=0.5in, bottom=1in]{geometry}
\pdfoutput=1
\usepackage{multirow}
\usepackage{array}
\usepackage{graphicx}
\usepackage{authblk}
\usepackage{booktabs}
\usepackage{amsmath}
\usepackage[table,xcdraw]{xcolor}
\usepackage{siunitx}
\usepackage[utf8]{inputenc}
\usepackage{geometry}
\usepackage{array}
\usepackage{graphicx}
\usepackage{tabularray}
\usepackage{multirow}
\usepackage{ragged2e}
\usepackage{float}
\usepackage{hyperref}
\usepackage{setspace}
\usepackage{subcaption}
\usepackage{booktabs}
\usepackage[title]{appendix}
\usepackage[flushleft]{threeparttable}
\usepackage{colortbl}
\usepackage{caption}

\providecommand{\keywords}[1]
{
  \small	
  \textbf{\textit{Keywords---}} #1
}

\title{Bayesian meta-analysis for evaluating treatment effectiveness in biomarker subgroups using trials of mixed patient populations}

\author[1]{Lorna Wheaton}
\author[2]{Dan Jackson}
\author[1]{Sylwia Bujkiewicz}

\date{}

\affil[1]{Biostatistics Research Group, Department of Population Health Sciences, University of Leicester, University Road, Leicester LE1 7RH, UK}

\affil[2]{Statistical Innovation Group, AstraZeneca, Cambridge, UK}

\newpage

\begin{document}

\maketitle

\section*{Abstract}
During drug development, evidence can emerge to suggest a treatment is more efficacious in a specific subgroup of patients. Whilst early trials may be conducted in biomarker-mixed populations, later trials are more likely to be conducted in biomarker-positive patients alone, thus leading to trials of the same treatment investigated in different populations. When conducting a meta-analysis, a conservative approach would be to combine only the trials conducted in the biomarker-positive subgroup. However, this discards potentially useful information on treatment effects in the biomarker-positive subgroup concealed within observed treatment effects in biomarker-mixed populations. We extend standard random-effects meta-analysis to combine treatment effects obtained from trials with different populations to estimate pooled treatment effects in the biomarker subgroup of interest. The model assumes a systematic difference in treatment effects between biomarker-positive and biomarker-negative subgroups, which is estimated from trials which report either or both treatment effects. The estimated systematic difference and proportion of biomarker-negative patients in biomarker-mixed studies are then used to interpolate treatment effects in the biomarker-positive subgroup from observed treatment effects in the biomarker-mixed population. The developed methods are applied to an illustrative example in metastatic colorectal cancer and evaluated in a simulation study. In the illustrative example, the developed method resulted in improved precision of the pooled treatment effect estimate compared to standard random-effects meta-analysis of trials investigating only biomarker-positive patients. The simulation study confirmed that when the systematic difference in treatment effects between biomarker subgroups is not very large, the developed method can improve precision of estimation of pooled treatment effects while maintaining low bias. 

\keywords{meta-analysis, biomarker, subgroup}

\newpage

\section{Introduction}

Traditionally, randomised controlled trials (RCTs) have aimed to obtain a reliable estimate of the average treatment effect in a broad patient population. However, over recent years there has been an increased interest in precision medicine where patient characteristics or biomarkers are used to characterise differences in disease risk, severity and efficacy of treatments and thus tailor healthcare to patients in order to maximise patient benefit~\cite{faulkner2020being}. Tailoring treatments to patients who are likely to benefit can also generate improvements in the cost-effectiveness of therapies~\cite{vijayaraghavan2012cost}.

During the course of drug development, evidence (for example from exploratory subgroup analyses) can emerge to suggest that a treatment is more effective in a particular subset of patients. When predictive biomarkers are used to identify subsets of a population which can benefit from treatments targeted on these biomarkers, they are investigated in clinical trials with mixed populations and trial designs~\cite{freidlin2010randomized, tajik2013trial}. For example, in metastatic colorectal cancer (mCRC), some novel therapeutic treatments were initially developed targeting the epidermal growth factor receptor (EGFR), overexpression of which (present in 50-80\% of colorectal tumours) can contribute to progression of the cancer~\cite{van2008kras}. However, evidence suggested that anti-EGFR treatments were only effective in a subgroup of patients and subsequent trials suggested that mutations in the kirsten rat sarcoma (KRAS) biomarker were predictive of resistance to EGFR-targeted treatments~\cite{saltz2004phase, van2016ras}. This resulted in early trials of anti-EGFR treatments investigating a mixture of patients with wild-type (WT) and mutant (MT) status of KRAS biomarker, while later trials only enrolled patients with KRAS WT status. Therefore, the development of EGFR-targeted therapies has resulted in trials with different designs and mixed populations.  

When assessing new therapies for their clinical and cost-effectiveness; for example, within a health technology assessment (HTA) framework, data from a range of trials are often combined in a meta-analysis. However, when RCTs are conducted in different populations, it is challenging to pool treatment effects in a single meta-analysis. One approach would be to only include studies which are conducted or report treatment effects in the subgroup of interest (e.g. biomarker-positive patients). However, this results in a loss of information (and therefore statistical power) as information on biomarker-positive patients contained within studies investigating patients with mixed biomarker status will not be utilised when subgroup results are unavailable. Individual participant data (IPD) would be required to conduct subgroup analyses of mixed populations where they were not reported, but it is unlikely that IPD could be obtained for all trials of mixed populations. An alternative approach is to include all studies in a single meta-analysis regardless of biomarker status. However, this will systematically increase between-trial heterogeneity and lead to an inconclusive treatment effect for the total population~\cite{thompson1994systematic}.

In this paper we modify the standard random-effects meta-analysis model to include treatment effects obtained from biomarker-positive, biomarker-negative and biomarker-mixed populations accounting for the differences in treatment effects across the biomarker groups in order to obtain a more precise estimate of the pooled treatment effect in the biomarker-positive subgroup. 

The remainder of this paper is structured as follows. An illustrative example in mCRC is described in Section \ref{section:illustrative_example} and the methods are described in Section \ref{section:methods}. Section \ref{section:illustrative_example_results} discusses the results from the application of the methods to the illustrative example. Section \ref{section:simulation} describes the methods and results of a simulation study used to evaluate the developed methods. Finally, discussion and conclusions are presented in Sections \ref{section:discussion} and \ref{section:conclusions}.

\section{Illustrative Example}\label{section:illustrative_example}

We use data from randomised controlled trials (RCTs) investigating anti-EGFR therapies for the treatment of mCRC. Over time, evidence has emerged suggesting that anti-EGFR therapies are effective in patients with WT KRAS biomarker status but ineffective in patients with MT KRAS biomarker~\cite{knickelbein2015mutant}. Anti-EGFR therapies have been evaluated in trials with different designs and with varying KRAS biomarker populations. 

\begin{figure}
    \centering
    \includegraphics[width=\textwidth]{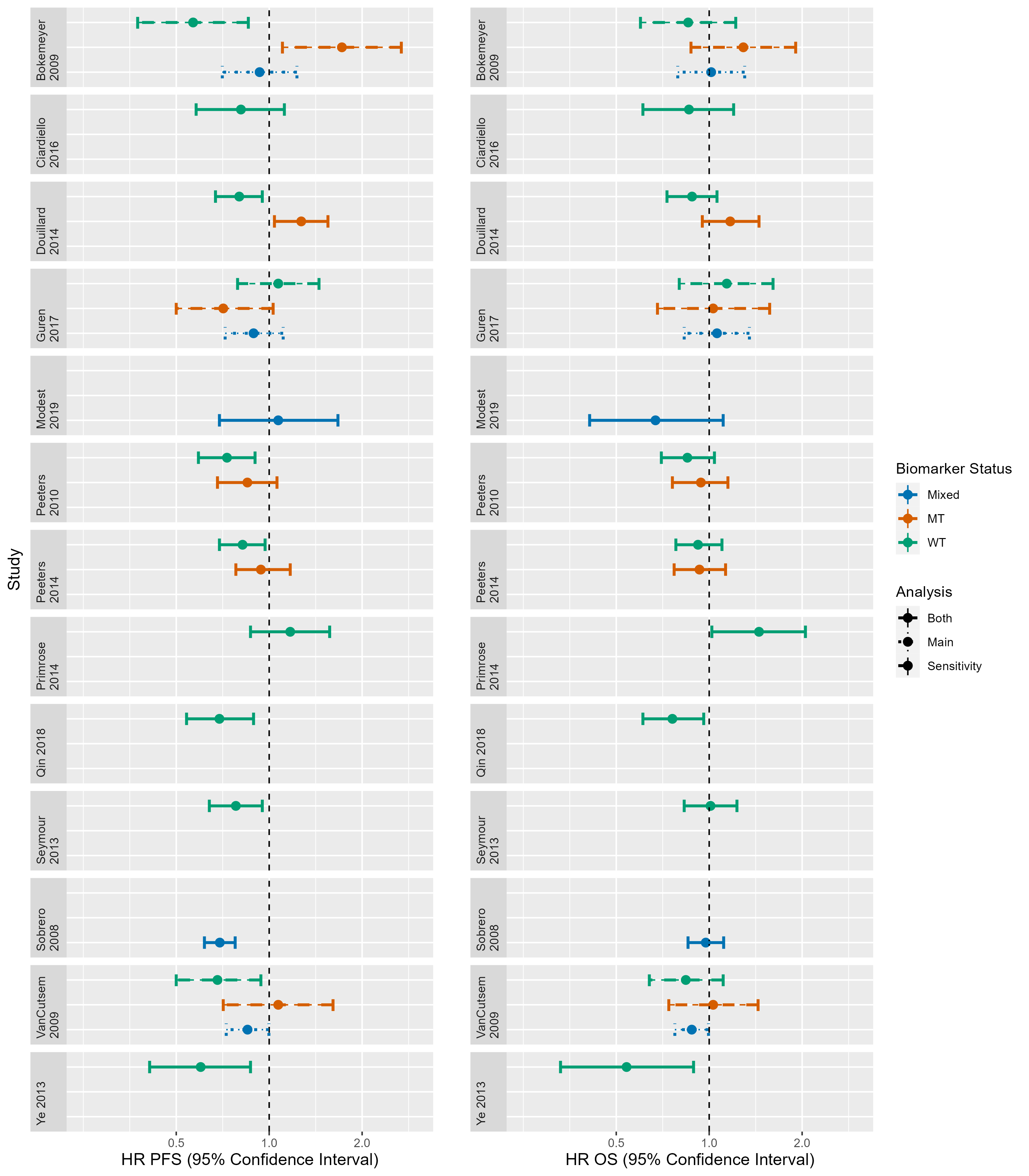}
    \caption{Observed hazard ratios and 95\% confidence intervals on progression-free survival and overall survival. Solid lines are used in both main and sensitivity analysis, dotted lines are used in main analysis only and dashed lines are used in sensitivity analysis only.}
    \label{fig:initialforest}
\end{figure}

The data used in the analysis were obtained from 13 RCTs. The studies reported hazard ratios (HRs) and corresponding confidence intervals (CIs) on progression-free survival (PFS) and overall survival (OS) which were converted to logHRs and their corresponding standard errors for use in analysis. Studies reported HRs and corresponding CIs for PFS and OS in at least one of the following groups; (1) patients with WT KRAS biomarker status, (2) patients with MT KRAS biomarker status and (3) patients with either WT or MT KRAS biomarker status (henceforth referred to as mixed). Of the 13 studies, 3 reported HRs for WT patients, MT patients and mixed patients, 3 reported HRs for WT and MT patients, 5 reported HRs for WT patients only and 2 reported HRs only for mixed patients (Figure \ref{fig:initialforest} \& Table \ref{table:alldata}). 

\begin{table}[]
\centering
\caption{LogHRs ($Y$) and corresponding standard errors ($\sigma$) on overall survival in KRAS WT ($+$), KRAS MT ($-$) and KRAS WT and MT ($mix$) populations from illustrative example in metastatic colorectal cancer used in main analysis. NA indicates no such treatment effect estimate is used from this study.}
\label{table:alldata}
\begin{tabular}{@{}lllllll@{}}
\toprule
Study                                                                   & $Y_{+}$ & $\sigma_{+}$ & $Y_{-}$ & $\sigma_{-}$ & $Y_{mix}$ & $\sigma_{mix}$ \\ \midrule
Bokemeyer 2009~\cite{bokemeyer2011efficacy}       & NA      & NA           & NA      & NA           & 0.01      & 0.13           \\
Ciardiello 2016~\cite{ciardiello2016cetuximab}    & -0.15   & 0.17         & NA      & NA           & NA        & NA             \\
Douillard 2014~\cite{douillard2014final}          & -0.13   & 0.10         & 0.16    & 0.11         & NA        & NA             \\
Guren 2017~\cite{tveit2012phase}                  & NA      & NA           & NA      & NA           & 0.06      & 0.12           \\
Modest 2019~\cite{swen2020folfoxiri}              & NA      & NA           & NA      & NA           & -0.40     & 0.25           \\
Peeters 2010~\cite{peeters2010randomized}         & -0.16   & 0.10         & -0.06   & 0.11         & NA        & NA             \\
Peeters 2014~\cite{peeters2014final}              & -0.08   & 0.09         & -0.07   & 0.10         & NA        & NA             \\
Primrose 2014~\cite{bridgewater2017perioperative} & 0.37    & 0.18         & NA      & NA           & NA        & NA             \\
Qin 2018~\cite{qin2018efficacy}                   & -0.27   & 0.12         & NA      & NA           & NA        & NA             \\
Seymour 2013~\cite{seymour2013panitumumab}        & 0.01    & 0.10         & NA      & NA           & NA        & NA             \\
Sobrero 2008~\cite{sobrero2008epic}               & NA      & NA           & NA      & NA           & -0.03     & 0.07           \\
Van Cutsem 2009~\cite{van2011cetuximab}           & NA      & NA           & NA      & NA           & -0.13     & 0.06           \\
Ye 2013~\cite{ye2013randomized}                   & 0.62    & 0.25         & NA      & NA           & NA        & NA             \\ \bottomrule
\end{tabular}
\end{table}

In the main analysis, where a study reported treatment effects from analysis of mixed patients in addition to reporting results from separate analyses of WT and MT patients (e.g. Bokemeyer 2009) only the treatment effect for the mixed population was included in the model (not using subgroup information). This resulted in 5 studies with mixed populations in the data and allowed us a better assessment of whether addition of results from the mixed biomarker population could improve estimation of pooled treatment effects for the WT group. The PFS and OS data used in the main analysis can be seen by the solid and dotted lines in Figure \ref{fig:initialforest}. In a sensitivity analysis, where a study reported treatment effects from analysis of mixed patients in addition to reporting results from separate analyses of WT and MT patients, the treatment effects for the WT and MT subgroup analyses were included in the model and the treatment effect from analysis of the mixed population was excluded (to avoid duplication of data). Inclusion of more information on the subgroups allowed us to assess the robustness of the developed methods. The PFS and OS data used in the sensitivity analysis can be seen by the solid and dashed lines in Figure \ref{fig:initialforest} respectively. 

To include studies from biomarker-mixed populations using the developed methods, the proportion of MT patients in the biomarker-mixed studies were required. For the studies Van Cutsem 2009, Guren 2017 and Bokemeyer 2009, KRAS status was known for 89\%, 86\% and 94\% of the total biomarker-mixed population giving proportions of KRAS MT patients of 0.37, 0.40 and 0.43 respectively. To account for uncertainty around these proportions, as not all patients have known KRAS status, we defined informative beta prior distributions for the proportion of KRAS MT patients using method of moments. However, for the studies Sobrero 2008 and Modest 2019, the proportion of MT patients was not reported. For these studies beta prior distributions were constructed using information on the prevalence of KRAS mutations suggesting that frequency of mutations was between 30\% and 54\%~\cite{neumann2009frequency}. This is consistent with proportions of KRAS MT patients in studies in this illustrative example which range from 0.37-0.45. Details of how the prior distributions were constructed are available in Appendix A.

\section{Methods}\label{section:methods}

To make the methods discussed in this paper generalisable to different scenarios, we shall henceforth refer to biomarker-positive, biomarker-negative and biomarker-mixed patient populations, where biomarker-positive patients are thought to respond to treatment and biomarker-negative patients are thought not to respond to treatment. Therefore, in the illustrative example in mCRC, patients with WT KRAS biomarker status are biomarker-positive and patients with MT KRAS biomarker status are biomarker-negative. In the following methods we are interested in estimating the pooled treatment effect in the biomarker-positive subgroup while utilising available information on treatment effects in the biomarker-negative and biomarker-mixed populations.

The model described in Section \ref{section:REMA_positive_only} below is a standard random-effects meta-analysis which we use to synthesise treatment effects from the biomarker-positive subgroup. In the illustrative example in mCRC, this model is applied to studies reporting HRs from analysis of patients with KRAS WT biomarker status only. The data utilised in the main analysis can be seen in the columns 2-3 of Table \ref{table:alldata}. In Section \ref{section:REMA_+-} the model is extended to include treatment effects from the biomarker-negative subgroup in order to estimate the systematic difference in treatment effects between the two subgroups. The model described in Section \ref{section:REMA_+-} is applied to studies reporting HRs from (1) analysis of patients with the KRAS WT biomarker only or (2) subgroup analysis of patients with KRAS WT and KRAS MT biomarkers. The data utilised in the main analysis can be seen in columns 2-5 of Table \ref{table:alldata}. Finally, in Section \ref{section:REMA+-mix} the model is extended to include treatment effects from the biomarker-mixed subgroup by utilising information on the systematic difference in treatment effects between biomarker subgroups, estimated in the second part of the model described in Section \ref{section:REMA_+-} and the proportion of biomarker-negative patients included in each biomarker-mixed study, in order to estimate treatment effects in the biomarker-positive subgroup. The model described in Section \ref{section:REMA+-mix} is applied to studies reporting HRs from (1) analysis of patients with the KRAS WT biomarker only, (2) subgroup analysis of patients with KRAS WT and KRAS MT biomarker or (3) analysis of mixed patients. The data utilised in the main analysis can be seen in Table \ref{table:alldata}. 

\subsection{Model 1: REMA for Biomarker-Positive Patients} \label{section:REMA_positive_only}

To estimate the pooled treatment effect in biomarker-positive patients, a standard approach would be to use random-effects meta-analysis (REMA) to combine treatment effects obtained from analysis of biomarker-positive patients only. In REMA, we assume that the normally distributed observed treatment effects in the biomarker-positive subgroup, $Y_{+i}$, are estimates of the true treatment effect, $\delta_{+i}$, in this group of patients: 

\begin{equation}\label{eq:withinWT}
    Y_{+i}|\delta_{+i} \sim N(\delta_{+i}, \sigma^{2}_{+i})
\end{equation}
with corresponding within-study variances, $\sigma_{+i}^2$, which are assumed to be known and equal to the squared observed standard error of the treatment effect estimate in each study $i=1,...,n_+$. The true effects are assumed to come from a normal distribution with a mean, $d_{+}$, and between-study variance, $\tau_{+}^2$:

\begin{equation}\label{eq:betweenWT}
    \delta_{+i} \sim N(d_{+}, \tau^{2}_{+})
\end{equation}

To implement this model using a Bayesian framework, the vague prior distributions $  d_{+} \sim N(0,100^2)$ and $ \tau_{+} \sim HN(0,10^2)$ (where HN indicates a half normal distribution) were placed on the pooled mean and between-study standard deviation. 

This model only uses information from studies which provide estimates of the treatment effect in biomarker-positive patients. In our illustrative example in mCRC, the model was applied to logHR data obtained from the analysis of patients with the KRAS WT biomarker status. The data for the main analysis are listed in columns 2-3 of Table \ref{table:alldata} and the data for the sensitivity analysis are listed in columns 2-3 of Table B1 in Appendix B. 

\subsection{Model 2: REMA for Biomarker-Positive and Biomarker-Negative Patients}\label{section:REMA_+-}

Bayesian REMA, described above, can be used to estimate pooled treatment effects for biomarker-positive patients when sufficient data exist. However, when there are a limited number of studies that report treatment effects in the biomarker-positive group such analysis can result in high uncertainty around the pooled estimate of the treatment effect. Therefore, we propose an extension to REMA to allow for inclusion of additional data obtained from the analysis of biomarker-negative patients. This can be achieved by assuming a systematic, albeit random, difference between the treatment effects in the two biomarker groups. 

The model for treatment effects from the analysis of biomarker-positive patients alone remains the same as described in Section \ref{section:REMA_positive_only}. For $i=n_{+} + 1,...,n_{+}+n_{\pm}$ where $n_{\pm}$ is the number of studies with effectiveness estimates available from both analysis of biomarker-positive and biomarker-negative patients, the treatment effects from analysis of biomarker-negative patients,  $Y_{-i}$, are assumed to be normally distributed with a true underlying mean $\delta_{-i}$, and within-study variance, $\sigma_{-i}^{2}$.

\begin{equation} \label{eqn:within_MT}
    Y_{-i}|\delta_{-i} \sim N(\delta_{-i}, \sigma^{2}_{-i})  
\end{equation}
The underlying true treatment effects in each study investigating biomarker-negative patients, $\delta_{-i}$, are then assumed to be equal to the underlying treatment effect estimated in biomarker-positive patients in study $i$, $\delta_{+i}$, plus a systematic difference, $\beta_{i}$: 

\begin{equation} \label{eqn:cond_delta_-}
    \delta_{-i} = \delta_{+i} + \beta_{i}
\end{equation}

The systematic difference between treatment effects in the biomarker-positive and biomarker-negative patients, $\beta_{i}$, can differ across studies and is assumed to come from a normal distribution with a mean, $\mu_{\beta}$, and variance, $\tau_{\beta}^2$:

\begin{equation} \label{eqn:between_beta1}
    \beta_{i} \sim N(\mu_{\beta}, \tau^{2}_{\beta}).
\end{equation}

To implement this model using a Bayesian framework, the vague prior distributions $ \tau_{\beta} \sim HN(0,10^2)$ and $\mu_{\beta} \sim N(0,100^2)$ were placed on the between-study standard deviation of the systematic difference and the mean of the systematic difference and $\delta_{+i}$ follow a common distribution as in equation (\ref{eq:betweenWT}) in Section \ref{section:REMA_positive_only}. The prior distributions for $d_{+}$ and $\tau_{+}$ remain the same as specified in Section \ref{section:REMA_positive_only}. 

This model uses information from studies which provide treatment effects from biomarker-positive patients only or both biomarker-positive and biomarker-negative patients. In the illustrative example in mCRC, the model was applied to data obtained from studies reporting results from analysis of KRAS WT patients only or studies reporting results from analysis of both the KRAS WT and KRAS MT biomarker subgroups. The data for the main analysis are listed in columns 2-5 of Table \ref{table:alldata} and the data for the sensitivity analysis are listed in columns 2-5 of Table B1 in Appendix B. 

It is worth acknowledging that this model can be used to include treatment effects from trials which only investigate biomarker-negative patients and details of this model are available in Appendix C. However, it is unlikely that studies will report treatment effects from analysis of biomarker-negative patients only as it is extremely unlikely that a clinical trial would be exclusively conducted in a subgroup where there is evidence that a treatment does not work. Therefore, studies which report treatment effects from biomarker-negative patients are also likely to report treatment effects from biomarker-positive patients and thus we focus on this scenario in this paper. 

\subsection{Model 3: REMA for Biomarker-Positive, Biomarker-Negative and Biomarker-Mixed patients}\label{section:REMA+-mix}

To include treatment effects from studies with a mix of biomarker-positive and biomarker-negative patients where there is no biomarker subgroup analysis reported, we extend the model further using information about the proportion of biomarker-negative patients in each study. The model for treatment effects from analysis of biomarker-positive and biomarker-negative patients remains the same as described in Section \ref{section:REMA_+-}. 

For $i=n_{+}+n_{\pm}+1,...,n_{+}+n_{\pm}+n_{mix}$, where $n_{mix}$ is the number of studies reporting only the analysis of data from biomarker-mixed patients, the observed treatment effects from analysis of biomarker-mixed patients, $Y_{mixi}$ are assumed normally distributed with an underlying mean, $\delta_{mixi}$ and within-study variance, $\sigma_{mixi}^{2}$. The underlying true treatment effects in each study $i$ investigating biomarker-mixed patients, $\delta_{mixi}$, are assumed to be equal to the underlying treatment effect for biomarker-positive patients in study $i$, $\delta_{+i}$, plus the systematic difference in treatment effects between biomarker-positive and biomarker-negative patients, $\beta_{i}$, multiplied by the proportion of biomarker-negative patients in the biomarker-mixed study, $p_{i}$.

\begin{equation}
    Y_{mixi}|\delta_{mixi} \sim N(\delta_{mixi}, \sigma^{2}_{mixi}) 
\end{equation}

\begin{equation} \label{eqn:assumption_of_linearity}
    \delta_{mixi} = \delta_{+i} + p_{i}\beta_{i} 
\end{equation}
In equation (\ref{eqn:assumption_of_linearity}) we assume that the treatment effect in the biomarker-mixed population has a linear relationship with the proportion of biomarker-negative patients in the study. It is important to acknowledge that the assumption of linearity is an approximation for outcomes such as logHRs. We consider this assumption further in the simulation study and discussion. For $p_{i}=0$ (biomarker-positive population) the model reduces to model M1 in Section \ref{section:REMA_positive_only} and for $p_{i}=1$ (biomarker-negative population) the model reduces to model M2 in Section \ref{section:REMA_+-}. As in the models described in Sections \ref{section:REMA_positive_only} \& \ref{section:REMA_+-}, the true treatment effects for biomarker-positive patients and systematic difference come from normal distributions with a mean and variance as specified in equations (\ref{eq:betweenWT}) \& (\ref{eqn:between_beta1}). To implement this model using a Bayesian framework the prior distributions described in Sections \ref{section:REMA_positive_only} and \ref{section:REMA_+-} were used. 

This model uses information from studies which provide treatment effects from analysis of biomarker-positive, biomarker-negative or biomarker-mixed populations. In the illustrative example in mCRC, the model was applied to logHR data obtained from studies reporting results from analysis of KRAS WT, KRAS MT or KRAS mixed biomarker groups. The data for the main analysis are listed in Table \ref{table:alldata} and the data for the sensitivity analysis are listed in Table B1 in Appendix B. 

In model M2, described in Section \ref{section:REMA_+-}, the biomarker-positive and biomarker-negative estimates can be assumed to be independent and thus model M2 can be applied using estimates from both subgroups. However, if a study were to provide treatment effect estimates from all groups (biomarker-positive, biomarker-negative and biomarker-mixed) these estimates would not be independent and would be highly correlated. Therefore, where a study reports all three treatment effects, it is not desirable for all three treatment effects to be included in model M3 (described here). Therefore, a decision must be made to either include (1) the subgroup estimates from biomarker-positive and biomarker-negative subgroups or (2) the estimates from the biomarker-mixed population only. In the main analysis of the illustrative example in this paper, where a study reported all three treatment effects the estimates from the biomarker-mixed population were included in model M3. This decision was made to more clearly illustrate whether inclusion of studies where only treatment effects on the biomarker-mixed population were available could improve precision of estimation of pooled treatment effects for the biomarker-positive subgroup. However, in a sensitivity analysis, where a study reported all three treatment effects the subgroup estimates were included in model M3. A comparison of results from model M3 in the main analysis and model M1 in the sensitivity analysis will give an indication of whether the developed model can reliably extract information on treatment effects in the biomarker-positive subgroup from observed treatment effects in a biomarker-mixed population in order to improve precision of estimation of pooled treatment effects in the biomarker-positive subgroup. 

\section{Results}\label{section:illustrative_example_results}

In this Section we report the results from applying the methods described in Section \ref{section:methods} to the example dataset described in Section \ref{section:illustrative_example}. REMA, described in Section \ref{section:REMA_positive_only}, was applied to treatment effects from analysis of patients with KRAS WT status only. In the main analysis, eight studies reported treatment effects (logHRs) on PFS and OS for the KRAS WT biomarker subgroup in the colorectal cancer applied example (solid green lines in Figure \ref{fig:initialforest}). In the sensitivity analysis, eleven studies reported treatment effects on PFS and OS for the KRAS WT biomarker subgroup (solid and dashed green lines in Figure \ref{fig:initialforest}). 

REMA extended to include treatment effects obtained from the biomarker-negative subgroup, described in Section \ref{section:REMA_+-}, was applied to treatment effects obtained from KRAS WT and KRAS MT biomarker subgroups separately. In the main analysis, five studies reported treatment effects on PFS and OS for the KRAS WT biomarker subgroup only and three studies reported treatment effects on PFS and OS for the KRAS WT and KRAS MT biomarker subgroups separately (solid green and red lines in Figure \ref{fig:initialforest}). In the sensitivity analysis, five studies reported treatment effects on PFS and OS for the KRAS WT biomarker subgroup only and six studies reported treatment effects on PFS and OS for KRAS WT and KRAS MT biomarker subgroups separately (solid and dashed green and red lines in Figure \ref{fig:initialforest}). 

REMA extended to include treatment effects obtained from biomarker-negative and biomarker-mixed populations, described in Section \ref{section:REMA+-mix}, was applied to treatment effects obtained from KRAS WT, KRAS MT and KRAS mixed populations. In the main analysis, in addition to the studies reporting treatment effects in KRAS WT and KRAS MT subgroups, five studies reported treatment effects in a mixed KRAS population (solid and dotted blue lines in Figure \ref{fig:initialforest}). In the sensitivity analysis, two studies reported treatment effects in a mixed KRAS population (solid blue lines in Figure \ref{fig:initialforest}).  

\begin{figure}[]

\subfloat[Progression-free survival]{%
  \includegraphics[clip,width=\columnwidth]{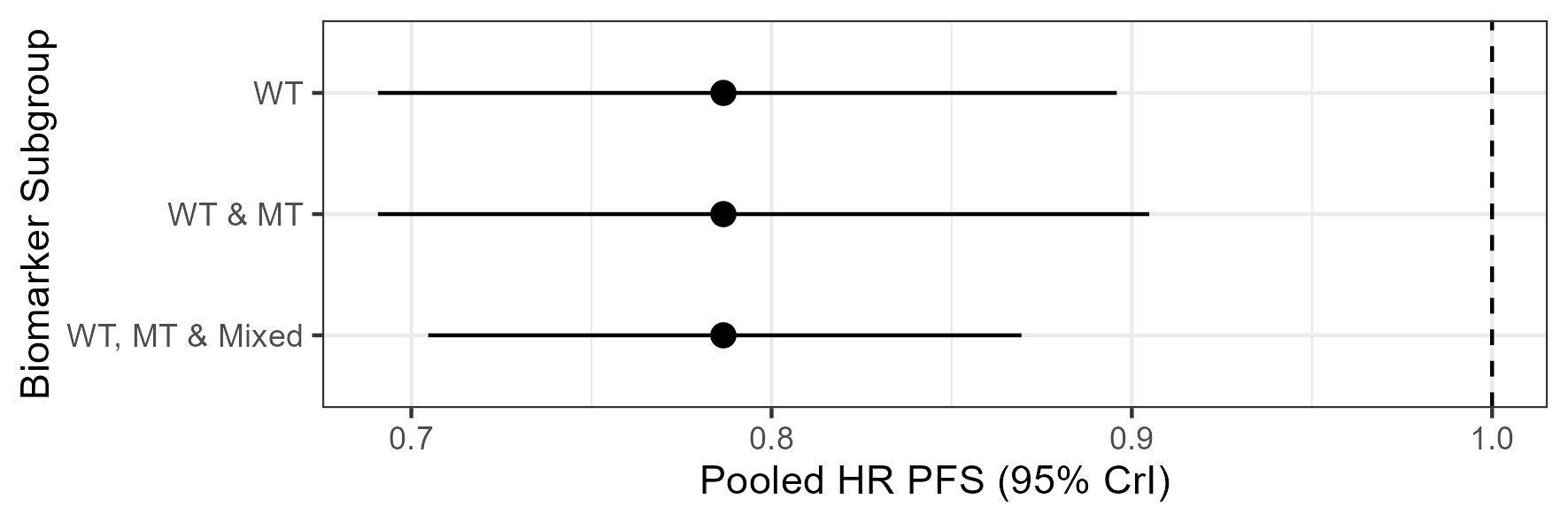}%
}

\subfloat[Overall survival]{%
  \includegraphics[clip,width=\columnwidth]{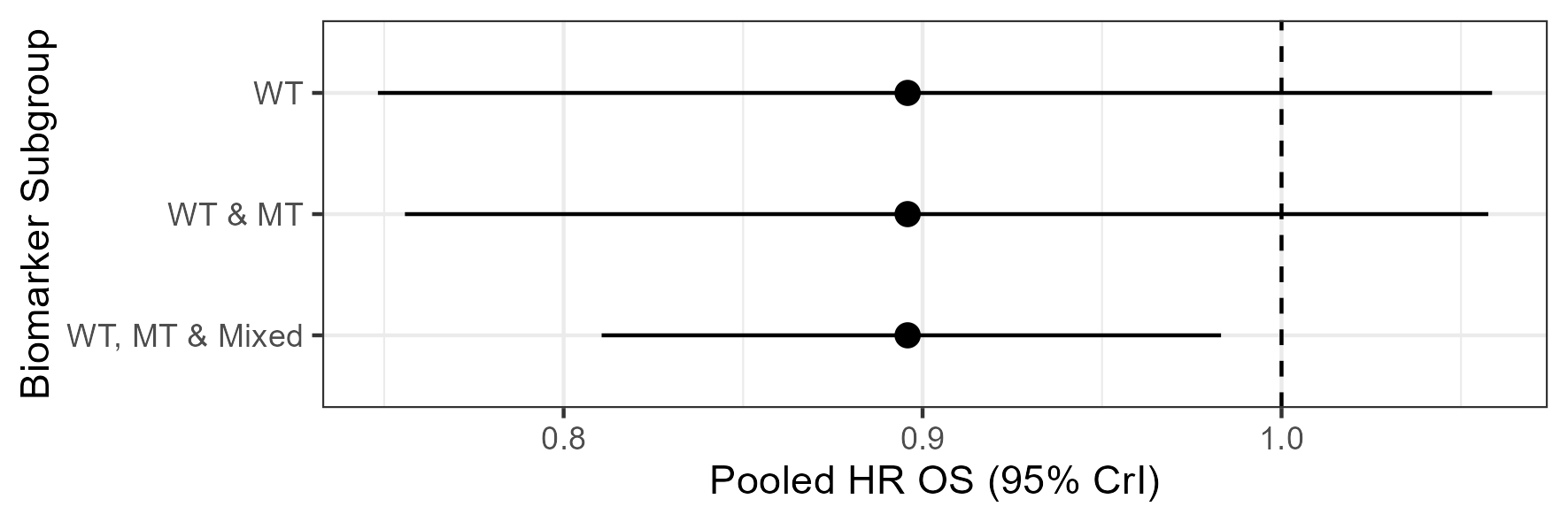}%
}

\caption{Main Analysis Results: Pooled treatment effects for PFS (a) and OS (b) for biomarker-positive subgroup estimated from models M1 (WT), M2 (WT \& MT) and M3 (WT, MT \& Mixed) in illustrative example in mCRC.}
\label{fig:pooledforestplot}

\end{figure}

Tables \ref{table:results} and \ref{table:sensitivityresults} show results from the main analysis and sensitivity analysis respectively. We consider the PFS outcome first. When only treatment effects from analysis of patients in the WT biomarker subgroup are used, the pooled logHR is estimated to be -0.24 (95\% CrI: -0.37, -0.11). This provides strong evidence for a meaningful treatment effect of anti-EGFR therapies in KRAS WT patients with mCRC as the point estimate is below zero and the 95\% CrI does not contain zero. The addition of treatment effects from analysis of patients with MT KRAS biomarker status to the meta-analysis slightly increases the CrI of the estimate of the pooled treatment effect for the biomarker-positive subgroup while the point estimate remains the same. However, the further addition of treatment effects from analysis of patients with mixed KRAS biomarker status estimates a pooled logHR for WT patients of -0.24 (95\% CrI: -0.35, -0.14). This indicates that addition of treatment effects from analysis of patients with mixed biomarker status, relative to using treatment effects from analysis of patients with WT biomarker status alone, results in a 19\% improvement in precision of estimation of pooled treatment effects for the WT biomarker subgroup. This improvement in precision is obtained without altering the point estimate of the pooled treatment effect. These results are illustrated in Figure \ref{fig:pooledforestplot}. 

\begin{table}
\caption{Main analysis results from application of models M1 (WT), M2 (WT \& MT) and M3 (WT, MT \& Mixed) to illustrative example in mCRC for the outcomes PFS and OS.}
\centering
\label{table:results}
\resizebox{\linewidth}{!}{%
\begin{tblr}{
  row{3} = {c},
  row{4} = {c},
  row{5} = {c},
  row{6} = {c},
  cell{1}{2} = {c=3}{c},
  cell{1}{5} = {c=3}{c},
  cell{2}{2} = {c},
  cell{2}{3} = {c},
  cell{2}{4} = {c},
  cell{2}{5} = {c},
  cell{2}{6} = {c},
  cell{2}{7} = {c},
  cell{3}{1} = {r},
  cell{4}{1} = {r},
  cell{5}{1} = {r},
  cell{6}{1} = {r},
  hline{1,7} = {-}{0.08em},
  hline{2} = {-}{0.05em},
  hline{3} = {-}{},
}
                 & PFS                        &                            &                             & OS                        &                           &                              \\
                 & WT                         & WT \& MT                     & WT, MT \& Mixed               & WT                        & WT \& MT                    & WT, MT \& Mixed                \\
$d_+$            & {-0.24 \\ (-0.37, -0.11)}  & {-0.24 \\ (-0.37, -0.10)}  & {-0.24 \\ (-0.35, -0.14)}   & {-0.11 \\ (-0.29, 0.057)} & {-0.11 \\ (-0.28, 0.056)} & {-0.11 \\ (-0.21, -0.017)}  \\
$\tau_+^2$       & {0.0076 \\ (0.0000, 0.11)} & {0.010 \\ (0.0000, 0.11)} & {0.0070 \\ (0.0000, 0.053)} & {0.020 \\ (0.0001, 0.21)} & {0.019 \\ (0.0001, 0.20)} & {0.0037 \\ (0.0000, 0.050)} \\
$\mu_{\beta}$    & NA                         & {0.25 \\ (-0.50, 1.00)}    & {0.22 \\ (-0.075, 0.51)}     & NA                        & {0.12 \\ (-0.48, 0.73)}   & {0.12 \\ (-0.094, 0.33)}      \\
$\tau_{\beta}^2$ & NA                         & {0.089 \\ (0.0004, 2.58)}   & {0.046 \\ (0.0003, 0.38)}   & NA                        & {0.036 \\ (0.0001, 2.11)}  & {0.010 \\ (0.0000, 0.18)}    
\end{tblr}
}
\end{table}

\begin{table}
\caption{Sensitivity analysis results from application of models M1 (WT), M2 (WT \& MT) and M3 (WT, MT \& Mixed) to illustrative example in mCRC for the outcomes PFS and OS.}
\centering
\label{table:sensitivityresults}
\resizebox{\linewidth}{!}{%
\begin{tblr}{
  row{3} = {c},
  row{4} = {c},
  row{5} = {c},
  row{6} = {c},
  cell{1}{2} = {c=3}{c},
  cell{1}{5} = {c=3}{c},
  cell{2}{2} = {c},
  cell{2}{3} = {c},
  cell{2}{4} = {c},
  cell{2}{5} = {c},
  cell{2}{6} = {c},
  cell{2}{7} = {c},
  cell{3}{1} = {r},
  cell{4}{1} = {r},
  cell{5}{1} = {r},
  cell{6}{1} = {r},
  hline{1,7} = {-}{0.08em},
  hline{2} = {-}{0.05em},
  hline{3} = {-}{},
}
                 & PFS                        &                             &                             & OS                          &                             &                              \\
                 & WT                         & WT \& MT                      & WT, MT \& Mixed               & WT                          & WT \& MT                      & WT, MT \& Mixed                \\
$d_+$            & {-0.24 \\ (-0.36, -0.12)}  & {-0.24 \\ (-0.34, -0.13)}   & {-0.25 \\ (-0.35, -0.14)}   & {-0.10 \\ (-0.22, 0.012)}   & {-0.10 \\ (-0.21, 0.0068)}  & {-0.11 \\ (-0.20, -0.014)}   \\
$\tau_+^2$       & {0.012 \\ (0.0001, 0.098)} & {0.0066 \\ (0.0000, 0.067)} & {0.0083 \\ (0.0000, 0.064)} & {0.0069 \\ (0.0001, 0.091)} & {0.0064 \\ (0.0000, 0.074)} & {0.0040 \\ (0.0000, 0.0052)} \\
$\mu_{\beta}$    & NA                         & {0.27 \\ (-0.11, 0.68)}     & {0.24 \\ (-0.10, 0.59)}     & NA                          & {0.14 \\ (-0.045, 0.34)}     & {0.13 \\ (-0.036, 0.31)}     \\
$\tau_{\beta}^2$ & NA                         & {0.10 \\ (0.0057, 0.99)}   & {0.12 \\ (0.0077, 0.73)}   & NA                          & {0.0086 \\ (0.0000, 0.17)}   & {0.0070 \\ (0.0000, 0.12)}   
\end{tblr}
}
\end{table}

We now consider the OS outcome. When only treatment effects from analysis of WT patients are used, the pooled logHR for WT patients is estimated to be -0.11 (95\% CrI: -0.29, 0.057). While the point estimate is below zero, the 95\% CrI contains zero indicating that for the OS outcome there is not strong evidence of a treatment effect for anti-EGFR therapies in the KRAS WT biomarker subgroup. As for the PFS outcome, the addition of treatment effects from analysis of MT patients does not change the results. However, the further addition of treatment effects from analysis of patients with mixed KRAS biomarker status estimates a pooled logHR for WT patients of -0.11 (95\% CrI: -0.21, -0.017). This provides stronger evidence for a treatment effect on OS of anti-EGFR therapies in the KRAS WT biomarker subgroup as the 95\% CrI is completely below zero. Addition of treatment effects from analysis of patients with mixed biomarker status, relative to using treatment effects from analysis of patients with WT biomarker status alone, results in a 44\% improvement in precision of estimation of pooled treatment effects for the WT biomarker subgroup. As for the PFS outcome, this improvement in precision is obtained without altering the point estimate of the pooled treatment effect. These results are illustrated in Figure \ref{fig:pooledforestplot}.

\begin{figure}[]

\subfloat[Progression-free survival]{%
  \includegraphics[clip,width=\columnwidth]{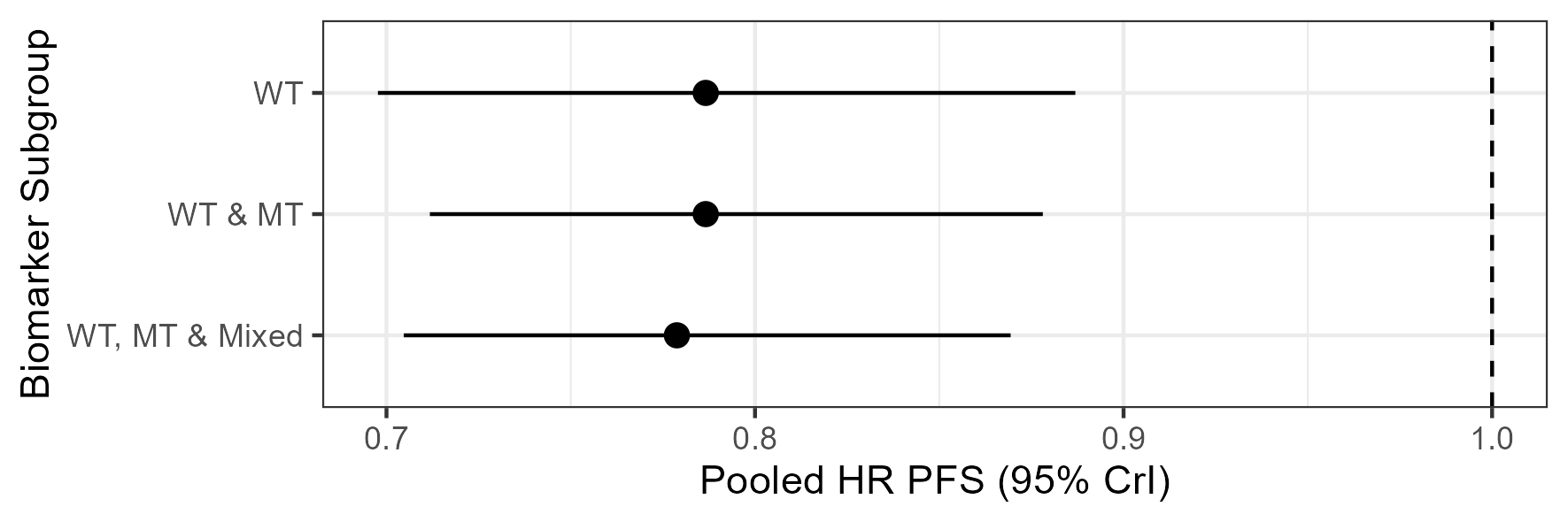}%
}

\subfloat[Overall survival]{%
  \includegraphics[clip,width=\columnwidth]{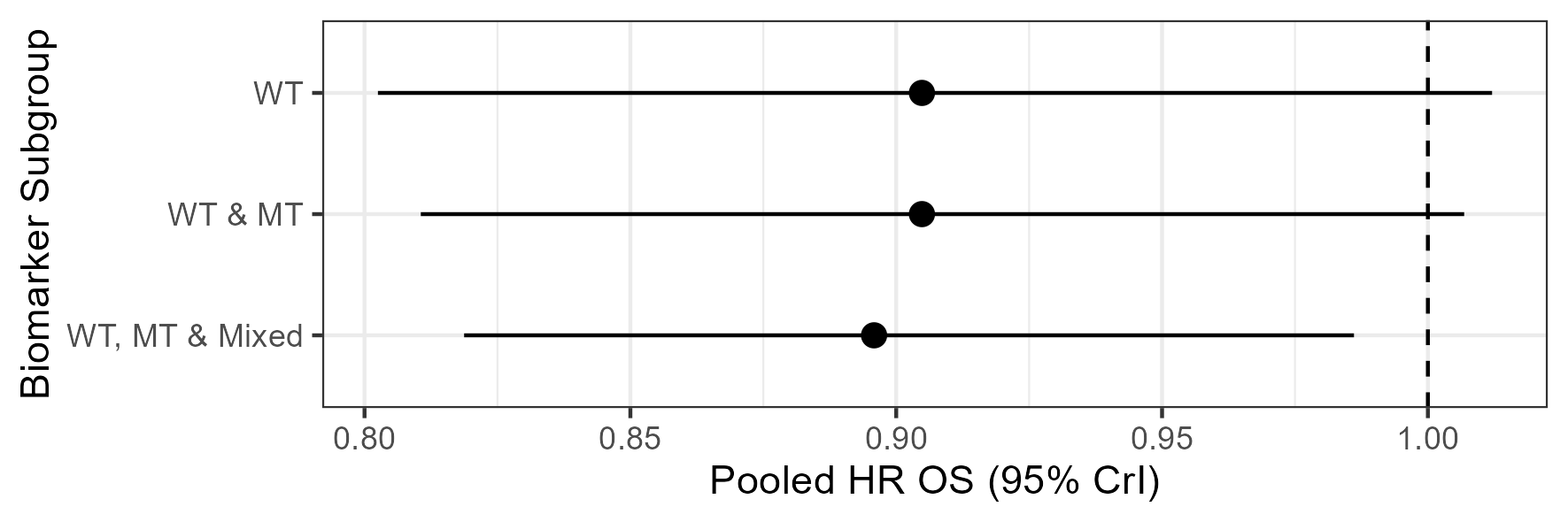}%
}

\caption{Sensitivity Analysis Results: Pooled treatment effects for PFS (a) and OS (b) for biomarker-positive subgroup estimated from models M1 (WT), M2 (WT \& MT) and M3 (WT, MT \& Mixed) in illustrative example in mCRC.}
\label{fig:pooledforestplot_sens}

\end{figure}

Pooled treatment effects on PFS and OS estimated by the sensitivity analysis utilising treatment effects from the biomarker-positive subgroup only were very similar to those estimated in the main analysis. Estimation of pooled treatment effects in the sensitivity analysis were generally more precise as a result of utilising data from subgroup analysis rather than mixed analysis. However, addition of treatment effects from analysis of biomarker-negative and biomarker-mixed groups still resulted in around 14\% and 20\% improvements in precision of estimation of the pooled treatment effect for the biomarker-positive subgroup on PFS and OS outcomes respectively, relative to using data on biomarker-positive patients alone. Potentially more importantly, the results from models M1 and M2 in the sensitivity analysis, where more subgroup data were available, are similar to results from model M3 in the main analysis, indicating robustness of the developed method. Results from the sensitivity analysis can be seen in Table \ref{table:sensitivityresults} and Figure \ref{fig:pooledforestplot_sens}. It is not surprising that improvements in precision are limited in the sensitivity analysis as only two additional studies reporting treatment effects in the biomarker-mixed population are available.

\section{Simulation Study}\label{section:simulation}

Results from the illustrative example suggest that the developed methods have the potential to improve precision of estimation of pooled treatment effects in the subgroup of interest. However, to formally assess whether the developed methods perform well and can improve upon standard methods, a simulation study is required. Therefore in this Section we report the methods and results of a simulation study designed to evaluate the performance of the developed methods. 

\subsection{Methods}

In this Section we report aims, data-generation methods, estimands, methods and performance measures for the simulation study, as recommended by Morris et al~\cite{morris2019using}. 

\subsubsection{Aims}

The simulation study aimed to compare the performance of the methods described in Section \ref{section:methods}, under a number of scenarios varying; (1) the proportion of studies reporting treatment effects on biomarker-positive and biomarker-negative subgroups, (2) the proportion of studies reporting treatment effects on the biomarker-mixed population, (3) the true mean systematic difference in treatment effects between biomarker subgroups, (4) the true variance of the systematic difference in treatment effects between biomarker subgroups and (5) the total number of studies in the meta-analysis. The data scenarios are discussed in more detail in Section \ref{section:datascenarios}. 

\subsubsection{Data-Generation Methods}

We set the total number of studies in each simulated meta-analysis to be $n_{studies}$. For $i=1,..., n_{studies}$ the true treatment effects in the biomarker-positive subgroup, $\delta_{+i}$, were drawn from a normal distribution with a mean, $d_{+}$, and variance, $\tau_{+}^2$ as in equation (\ref{eq:betweenWT}). For $i=1,...,n_{studies}$ the true systematic difference in treatment effects between biomarker-positive and biomarker-negative subgroups, $\beta_{i}$, was drawn from a normal distribution with a mean, $\mu_{\beta}$, and variance $\tau_{\beta}^2$ as in equation (\ref{eqn:between_beta1}). The true treatment effects in the biomarker-negative subgroup, $\delta_{-i}$, were then assumed to be equal to the sum of the true treatment effects in the biomarker-positive subgroup and the systematic difference as in equation (\ref{eqn:cond_delta_-}). 

Having generated true treatment effects for each study $i=1,..,n_{studies}$, in the meta-analysis, IPD were generated separately for biomarker-positive and biomarker-negative subgroups using the \textbf{simsurv} command in R~\cite{brilleman2021simulating}. To do this we set the number of participants in each individual trial $n_{participants}=350$, the probability of receiving treatment, $p_{trt}=0.5$ and the probability of having negative biomarker status, $p_{-} \sim Beta(9.2, 13.8)$. The probability of being biomarker-negative was based on the prevalence of KRAS mutations in mCRC~\cite{neumann2009frequency}. 

We assumed that survival data for the biomarker-positive and biomarker-negative subgroups in each study $i$ were drawn from two separate exponential distributions each with a single rate parameter $\lambda_{+}$ and $\lambda_{-}$ defining the baseline hazard. We assumed that $\lambda=\lambda_{+}=\lambda_{-}$, that is, that the baseline hazard (that is the hazard in the control arm) was the same for biomarker-positive and biomarker-negative subgroups. We set $\lambda=0.15$ as this was the baseline hazard estimated from analysis of the illustrative example in mCRC.

Survival data for the biomarker-positive subgroup were drawn from an exponential distribution with a rate parameter of $\lambda$ and an assumed treatment effect of $\delta_{+i}$. Survival data for the biomarker-negative subgroup were drawn from an exponential distribution with a rate parameter $\lambda_{-}$ and an assumed treatment effect of $\delta_{-j}$. This resulted in the hazard rates for the treatment and control arms in the biomarker-positive and biomarker-negative subgroups as shown in Table \ref{table:hazardrates}. 

\begin{table}[H]
\centering
\caption{Hazard rates for control and treatment arms in biomarker-positive and biomarker-negative subgroups for study $i$ in simulation study.}
\label{table:hazardrates}
\begin{tabular}{lll}
                   & Control        & Treatment                                                     \\
Biomarker-positive & $Exp(\lambda)$ & $Exp(\lambda+\delta_{+i})$                                    \\
Biomarker-negative & $Exp(\lambda)$ & $Exp(\lambda+\delta_{-i})$
\end{tabular}
\end{table}

Individual participant data for the biomarker-mixed population in study $i$ were simply created by combining the survival times from the biomarker-positive and biomarker-negative subgroups into a single dataset. 

For each study $i$ in each simulated meta-analysis, the biomarker-positive and biomarker-negative IPD were analysed by subgroup using the Cox proportional hazards model with a single covariate for treatment arm which can be expressed using the following hazard function:  $h_{j}(t)=h_{0}(t) \times exp(b_{1}trt_{j})$, where $trt_{j}$ is a binary variable for treatment group ($trt_{j}=0$ indicating control and $trt_{j}=1$ indicating treatment). 

The biomarker-mixed IPD were analysed in two ways; the first was using the Cox proportional hazards model with a single covariate for treatment arm (as in the equation above) and the second was using a Cox proportional hazards model with a covariate for treatment and a covariate for biomarker status which can be expressed using the following hazard function: $h_{j}(t)=h_{0}(t) \times exp(b_{1}trt_{j} + b_{2}biomarker_{j})$, where $biomarker_{j}$ is a binary variable for biomarker status ($biomarker_{j}=0$ indicating biomarker-positive status and $biomarker_{j}=1$ indicating biomarker-negative status). 

This resulted in four treatment effects being estimated for each study $i$ in each simulated meta-analysis. These were; treatment effect in the biomarker-positive subgroup, treatment effect in the biomarker-negative subgroup, treatment effect in biomarker-mixed population and treatment effect adjusted for biomarker status in the biomarker-mixed population. The treatment effects and associated standard errors estimated by the Cox model were saved as the observed treatment effects, $Y_{+i}$, $,Y_{-i}$, $Y_{mixi}$ and $Y_{Adjmixi}$ and within-study standard deviations, $\sigma_{+i}$, $\sigma_{-i}$, $\sigma_{mixi}$ and $\sigma_{Adjmixi}$. 

To replicate the illustrative example where treatment effects are not available from biomarker-positive, biomarker-negative and biomarker-mixed groups in every study, observed treatment effects are assumed to be missing for some of the $n_{studies}$ in the meta-analysis. We assume there are $n_{studies+}$ reporting treatment effects in the biomarker-positive subgroup alone, $n_{studies\pm}$ reporting treatment effects in biomarker-positive and biomarker-negative subgroups and $n_{studiesmix}$ reporting treatment effects in the biomarker-mixed population. To ensure no double counting of data, we assume that studies which report treatment effects in the biomarker-mixed population do not report treatment effects in the biomarker-positive or biomarker-negative subgroups.

\subsubsection{Estimand}

The estimand of interest is the pooled treatment effect in the biomarker-positive subgroup. 

\subsubsection{Methods}

Model M1 was applied to treatment effects from analysis of biomarker-positive patients only. Model M2 was applied to treatment effects from subgroup analysis of biomarker-positive and biomarker-negative patients. Model M3 was applied to treatment effects from analysis of biomarker-positive, biomarker-negative and biomarker-mixed populations. However, in the simulation, estimated treatment effects for the biomarker-mixed population could be obtained from a Cox proportional hazards model without adjustment for biomarker status or with adjustment for biomarker status. Therefore, in the following, application of model M3 to treatment effects from the biomarker-mixed population unadjusted for biomarker status will be referred to as model M3-unadj and application of model M3 to treatment effects from the biomarker-mixed population adjusted for biomarker status will be referred to as M3-adj. It is important to note that the only difference between M3-unadj and M3-adj are the estimated treatment effects for the biomarker-mixed population which are used as inputs for model M3.

\subsubsection{Performance Measures}

We evaluated performance by calculating the percentage bias, coverage and mean width of the credible interval for the pooled treatment effect in the biomarker-positive subgroup. The methods were implemented via MCMC sampling in the WinBUGS software, using a burn-in of 50,000 iterations and 100,000 iterations for posterior estimation ~\cite{lunn2000winbugs}. 

\subsubsection{Data Scenarios}\label{section:datascenarios}

The data were simulated under a number of scenarios adapted from scenario 1 (S1), where the number of studies in the meta-analysis was $n_{studies}=15$, the number of studies reporting observed treatment effects in the biomarker-positive subgroup only was $n_{studies+}=5$, the number of studies reporting observed treatment effects in the biomarker-positive and biomarker-negative subgroup was $n_{studies\pm}=5$, the number of studies reporting treatment effects in the biomarker-mixed group was $n_{studiesmix}=5$, the true pooled treatment effect in the biomarker-positive subgroup was $d_{+}=-0.25$, the true between-study variance in the biomarker-positive subgroup was $\tau_{+}^2=0.0056$, the true mean of the systematic difference was $\mu_{\beta}=0.25$ and the true variance of the systematic difference was $\tau_{\beta}^2=0.01$. 

We arrange the scenarios into five groups, where in each group the scenarios focus on varying a common set of parameter values. In group one, [S1-S5], the number of trials reporting observed treatment effects from the biomarker-positive subgroup only decreases (from $n_{studies+}=5$ to $n_{studies+}=1$) as the number of trials reporting observed treatment effects from biomarker-positive and biomarker-negative subgroups increases (from $n_{studies\pm}=5$ to $n_{studies\pm}=9$). This was intended to clearly demonstrate the performance of the methods as the number of studies used to estimate the systematic difference increases (i.e. S4 and S5 where 80\% and 90\% of studies reporting biomarker-positive treatment effects also report biomarker-negative treatment effects relative to S1 where only 50\% of studies reporting biomarker-positive treatment effects also report biomarker-negative treatment effects). 

In group two [S1, S6-S9], the number of treatment effects from biomarker-positive analysis only gradually decreases (from $n_{studies+}=5$ to $n_{studies+}=1$) and the number of treatment effects from biomarker-mixed groups gradually increases (from $n_{studiesmix}=5$ to $n_{studiesmix}=9$). This was intended to clearly demonstrate the performance of the methods as the number of studies with biomarker-mixed analysis, relative to the number of studies with biomarker-positive analysis increases. 

In group three, [S1, S10-S13] the mean of the systematic difference increases (from $\mu_{\beta}=0.25$ to $\mu_{\beta}=1.25$). This was intended to clearly demonstrate the performance of the methods as the difference between the true treatment effects observed in the biomarker-positive and biomarker-negative subgroups increases. 

In group four, [S1, S14-S17] the variance of the systematic difference gradually increases (from $\tau_{\beta}^2=0.01$ to $\tau_{\beta}^2=0.3$). This was intended to clearly demonstrate the performance of the methods as the variability of the systematic difference in the treatment effects between biomarker-positive and biomarker-negative subgroups increases. 

In group five, [S1, S18-S21] the total number of studies in the meta-analysis increases (from $n_{studies}=9$ to $n_{studies}=90$). This was intended to clearly demonstrate the performance of the methods as the availability of data increases. 

For all scenarios the true pooled treatment effect and between-study variance for the biomarker-positive subgroup were set as $d_{+}=0.25$ and $\tau_{+}^2=0.0056$. A full description of the parameter values specified in each scenario is provided in Table \ref{table:scenarios}.

\begin{table}[]
\centering
\caption{Model parameters specified to simulate datasets under 21 scenarios}
\label{table:scenarios}
\begin{tabular}{@{}llllllllll@{}}
\toprule
Scenario & Group & $n_{studies}$ & $n_{studies+}$ & $n_{studies\pm}$ & $n_{studiesmix}$ & $\mu_{\beta}$ & $\tau_{\beta}^2$ & $d_{+}$ & $\tau_{+}^2$ \\ \midrule
S1       & 1-5   & 15            & 5              & 5               & 5                  & 0.25          & 0.01             & -0.25   & 0.0056       \\
S2       & 1     & 15            & 4              & 6               & 5                  & 0.25          & 0.01             & -0.25   & 0.0056       \\
S3       & 1     & 15            & 3              & 7               & 5                  & 0.25          & 0.01             & -0.25   & 0.0056       \\
S4       & 1     & 15            & 2              & 8               & 5                  & 0.25          & 0.01             & -0.25   & 0.0056       \\
S5       & 1     & 15            & 1              & 9               & 5                  & 0.25          & 0.01             & -0.25   & 0.0056       \\
S6       & 2     & 15            & 4              & 5               & 6                  & 0.25          & 0.01             & -0.25   & 0.0056       \\
S7       & 2     & 15            & 3              & 5               & 7                  & 0.25          & 0.01             & -0.25   & 0.0056       \\
S8       & 2     & 15            & 2              & 5               & 8                  & 0.25          & 0.01             & -0.25   & 0.0056       \\
S9       & 2     & 15            & 1              & 5               & 9                  & 0.25          & 0.01             & -0.25   & 0.0056       \\
S10      & 3     & 15            & 5              & 5               & 5                  & 0.5           & 0.01             & -0.25   & 0.0056       \\
S11      & 3     & 15            & 5              & 5               & 5                  & 0.75          & 0.01             & -0.25   & 0.0056       \\
S12      & 3     & 15            & 5              & 5               & 5                  & 1             & 0.01             & -0.25   & 0.0056       \\
S13      & 3     & 15            & 5              & 5               & 5                  & 1.25          & 0.01             & -0.25   & 0.0056       \\
S14      & 4     & 15            & 5              & 5               & 5                  & 0.25          & 0.05             & -0.25   & 0.0056       \\
S15      & 4     & 15            & 5              & 5               & 5                  & 0.25          & 0.1              & -0.25   & 0.0056       \\
S16      & 4     & 15            & 5              & 5               & 5                  & 0.25          & 0.2              & -0.25   & 0.0056       \\
S17      & 4     & 15            & 5              & 5               & 5                  & 0.25          & 0.3              & -0.25   & 0.0056       \\
S18      & 5     & 9             & 3              & 3               & 3                  & 0.25          & 0.01             & -0.25   & 0.0056       \\
S19      & 5     & 30            & 10             & 10              & 10                 & 0.25          & 0.01             & -0.25   & 0.0056       \\
S20      & 5     & 60            & 20             & 20              & 20                 & 0.25          & 0.01             & -0.25   & 0.0056       \\
S21      & 5     & 90            & 30             & 30              & 30                 & 0.25          & 0.01             & -0.25   & 0.0056       \\ \bottomrule
\end{tabular}
\end{table}

\subsection{Results}

In this Section, we present the simulation study results for each method in terms of percentage bias, coverage and mean width of the credible interval. Here the methods described in Sections \ref{section:REMA_positive_only} and \ref{section:REMA_+-} are referred to as methods M1 and M2 respectively. The method described in Section \ref{section:REMA+-mix} utilising treatment effects from biomarker-mixed populations without adjustment for biomarker group is referred to as model M3-unadj and the same method utilising treatment effects from biomarker-mixed populations adjusted for treatment group and biomarker status is referred to as model M3-adj. We illustrate the results for each scenario group with a line plot of the performance measures for the estimand.

\subsubsection*{Scenarios S1-S5}

Across scenarios [S1-S5], the proportion of studies reporting observed treatment effects in both biomarker-positive and biomarker-negative subgroups increases whilst the proportion of studies reporting observed treatment effects in the biomarker-positive subgroup alone decreases. 

The results for these scenarios are presented in Figure \ref{fig:lineplot_s1_s9}. There is reasonably small percentage bias (below 1.5\%) and over-coverage (i.e. coverage above the nominal value 0.95) for all models in all scenarios. However, for models M3-unadj and M3-adj, which include treatment effects from biomarker-positive, biomarker-negative and biomarker-mixed populations, the mean CrI is much lower than for models M1 and M2 which only include treatment effects from biomarker-positive and biomarker-positive and biomarker-negative subgroups respectively. This indicates that inclusion of treatment effects from biomarker-mixed populations informed by the systematic difference in treatment effects between biomarker-positive and biomarker-negative subgroups, improves the precision of estimation of pooled treatment effects for the biomarker-positive group of interest. For all performance measures and for all models it does not appear that changing the proportion of studies reporting treatment effects from the biomarker-negative subgroup, impacts performance.

\subsubsection*{Scenarios S1 \& S6-S9}

Across scenarios [S1, S6-S9], the proportion of studies reporting observed treatment effects from the biomarker-positive subgroup decreases (from $n_{studies+}=5$ to $n_{studies+}=1$) and the proportion of studies reporting observed treatment effects from the biomarker-mixed population increases (from $n_{studiesmix}=5$ to $n_{studiesmix}=9$). 

\begin{figure}[]
  \includegraphics[width=\linewidth]{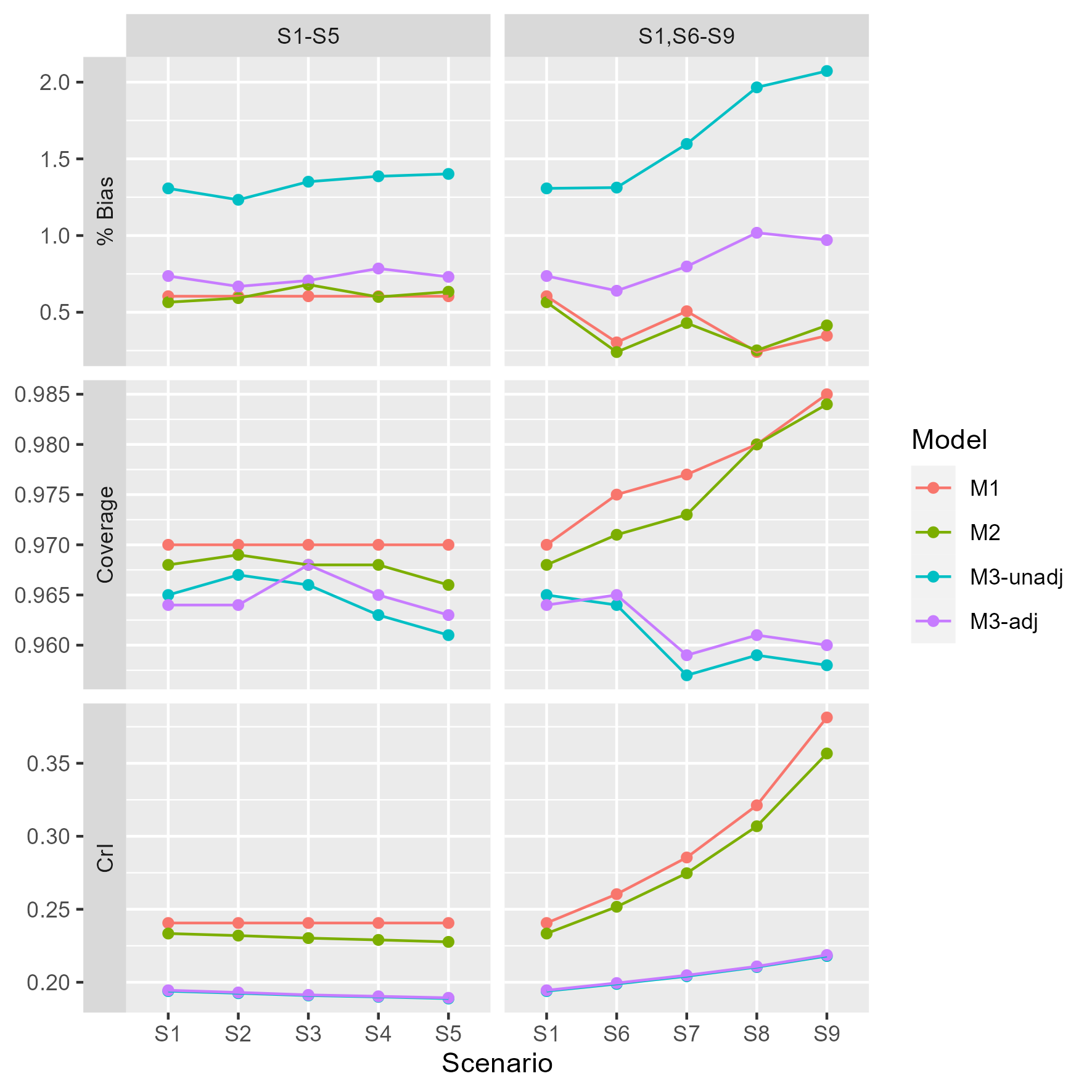}
  \caption{Percentage bias, coverage and mean credible interval width for pooled treatment effects for the biomarker-positive subgroup, across scenarios; (a) S1-S5 where the proportion of studies reporting treatment effects in the biomarker-negative subgroup in addition to treatment effects in the biomarker-positive subgroup increases and (b) S1 \& S6-S9 where the proportion of studies reporting observed treatment effects from the biomarker-positive subgroup decreases and the proportion of studies reporting observed treatment effects from the biomarker-mixed population increases.} 
  \label{fig:lineplot_s1_s9}
\end{figure}

The results for these scenarios are presented in Figure \ref{fig:lineplot_s1_s9}. As for scenarios [S1-S5], for all models the percentage bias is relatively small, there is over-coverage and models M3-unadj and M3-adj achieve lower uncertainty than models M1 and M2 in all scenarios. However, for scenarios [S1, S6-S9], the mean width of the 95\% CrI increases as the number of studies reporting treatment effects from the biomarker-positive subgroup decreases and the number of studies reporting treatment effects from the biomarker-mixed population increases. This is to be expected as when there are fewer treatment effects available from the biomarker-positive subgroup 
there is greater uncertainty around the estimate of the pooled treatment effect for this subgroup. However, the ability to utilise treatment effect estimates on the biomarker-mixed subgroup when using models M3-unadj and M3-adj results in a shallower increase in uncertainty around estimates of the pooled treatment effect compared to using treatment effects from the biomarker-positive subgroup alone. 

\subsubsection*{Scenarios S1 \& S10-S13}

Across scenarios [S1, S10-S13], the true mean of the systematic difference gradually increased (from $\mu_{\beta}=0.25$ to $\mu_{\beta}=1.25$). 

The results for these scenarios are presented in Figure \ref{fig:lineplot_s1_s10_s17}. Percentage bias is similar and close to zero for models M1, M2 and M3-adj. As in the previous groups, models M3-unadj and M3-adj result in more precise estimates than models M1 and M2. However, for M3-unadj the percentage bias increases as the true mean of the systematic difference in treatment effects between biomarker-positive and biomarker-negative subgroups increases. This is likely due to non-collapsibility of the hazard ratio. However, including treatment effects from the biomarker-mixed population adjusted for biomarker status in model M3-adj results in only a minimal increase in the percentage bias as the systematic difference increases. Thus, using treatment effects from biomarker-mixed populations adjusted for biomarker status moderates the effect of the increase in systematic difference. 

\subsubsection*{Scenarios S1 \& S14-S17}

Across scenarios [S1, S14-S17], the variance of the systematic difference in treatment effects between biomarker-positive and biomarker-negative subgroups gradually increased (from $\tau_{\beta}^2=0.01$ to $\tau_{\beta}^2=0.3$). 

\begin{figure}[]
  \includegraphics[width=\linewidth]{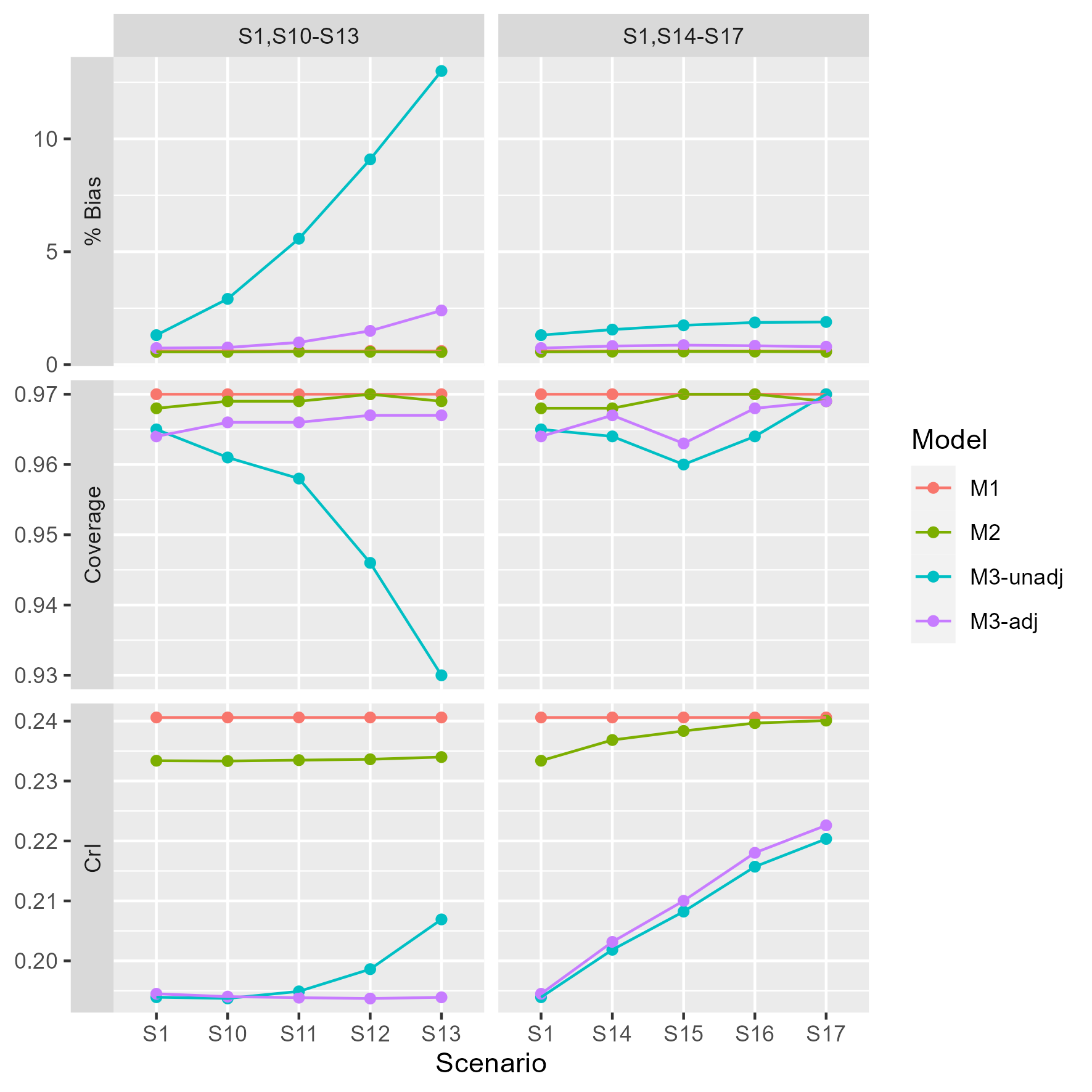}
  \caption{Percentage bias, coverage and mean credible interval width for pooled treatment effects for the biomarker-positive subgroup, across scenarios; (a) S1, \& S10-S13 where the true mean systematic difference between treatment effects in the biomarker-positive and biomarker-negative group gradually increases and (b) S1 \& S14-S17 where the true variance of the systematic difference between treatment effects in the biomarker-positive and biomarker-negative group gradually increases.} 
  \label{fig:lineplot_s1_s10_s17}
\end{figure}

The results for these scenarios are presented in Figure \ref{fig:lineplot_s1_s10_s17}. Percentage bias is small and similar for models M1, M2 and M3-adj while the percentage bias is larger for M3-unadj and increases as the variance of the systematic difference increases. Once again, there is over-coverage for all models. Models M3-unadj and M3-adj including treatment effects from biomarker-mixed populations achieve lower CrIs than models M1 and M2 which do not utilise treatment effects from the biomarker-mixed population. However, as the variance of the systematic difference increases the improvement in precision when using models M3-unadj and M3-adj compared to models M1 and M2 decreases.

\subsubsection*{Scenarios S1 \& S18-S21}

Across scenarios [S1, S18-S21], the total number of studies in the meta-analysis increases (from $n_{studies}=9$ to $n_{studies}=90$). 

The results for these scenarios are presented in Figure D1 in Appendix D. For all four models percentage bias and coverage are generally similar and decreasing as the number of studies in the meta-analysis increases. As for all other simulation scenarios, the mean CrI is always lower when using models M3-unadj and M3-adj compared to models M1 and M2. However, as the number of studies in the meta-analysis increases, there is less potential to improve the precision of pooled treatment effect estimates by utilising additional data from biomarker-negative and biomarker-mixed populations.

\section{Discussion}\label{section:discussion}

In this paper, we describe methods for synthesis of data on treatment effects from studies with varying reporting of biomarker subgroup analyses in order to estimate pooled treatment effects for a biomarker subgroup of interest. We applied these methods to an illustrative example in mCRC, where a proportion of studies did not report subgroup analysis. This resulted in a 19\% and 44\% improvement in precision of estimates of the pooled treatment effects on PFS and OS respectively compared to the standard method applied to data available from subgroup analyses only. These results were consistent with results from a sensitivity analysis utilising more subgroup information. This indicated that the addition of treatment effects from KRAS MT and KRAS mixed populations improved precision of estimates of pooled treatment effects in the KRAS WT population without introducing bias.

To formally assess whether the model can improve precision without introducing bias, we carried out a simulation study. Based on the simulation study, we conclude that model M3-unadj provides a consistent reduction in uncertainty compared to models M1 and M2. However, this reduction in uncertainty is accompanied by a consistent increase in percentage bias for model M3-unadj compared to models M1 and M2. Furthermore for model M3-unadj, unlike models M1 and M2, the percentage bias obtained increases as (a) the true systematic difference between treatment effects in the two biomarker subgroups increases and (b) the variance of the systematic difference increases. It is not surprising that the bias increases as the true systematic difference in treatment effects between biomarker-positive and biomarker-negative subgroups increases. This is because model M3-unadj assumes that the treatment effect in the biomarker-mixed population is linear with the proportion of biomarker-negative patients in the study and this assumption of linearity is known not to hold when the effect size is non-collapsible. However, in data scenarios where the true systematic difference is low, the linearity assumption made in model M3-unadj has the potential to improve precision without inducing any significant bias on the estimated pooled treatment effect in the biomarker-positive subgroup. 

Moreover, the model M3-adj (where treatment effects from biomarker-mixed populations are adjusted for biomarker status) improves on model M3-unadj by consistently achieving lower uncertainty and percentage bias than model M3-unadj. Furthermore, model M3-adj is reasonably robust to (a) increases in the systematic difference and (b) increases in the variance of the systematic difference. While percentage bias does increase for model M3-adj across scenarios [S1, S10-S13], percentage bias only increases significantly above that obtained by models M1 and M2 in scenarios S12 and S13 where the true systematic difference is very large and unlikely to be seen in practice. For example, in scenario S12 the HR in the biomarker-positive subgroup is 0.77 and the HR in the biomarker-negative subgroup is 2.12. This suggests that in cases where the true systematic difference is large, but still plausible, model M3-adj has the potential to improve precision without inducing any significant bias on the estimated pooled treatment effect in the biomarker-positive subgroup. Therefore, in data scenarios where the systematic difference between treatment effects in the subgroups is likely to be large, it is preferable to apply model M3 to treatment effects from biomarker-mixed populations which are adjusted for treatment group and biomarker status. However, we acknowledge that in practice, studies conducted in biomarker-mixed populations are unlikely to report treatment effects adjusted for biomarker status. 

There are some limitations of the analyses and methods described in this paper. First, in the meta-analysis model it is assumed that the logHR treatment effects are normally distributed, which is a strong assumption. Second, the within-study variances which are used as inputs to the meta-analysis models are obtained from standard errors of analysis of the IPD using the Cox proportional hazards model. Therefore standard error estimates will not be accounting for the random-effects structure of the data. Finally, it should be acknowledged that the assumption of linearity of treatment effects in the biomarker-mixed population with the proportion of biomarker-negative patients in a study is a strong assumption which will not hold exactly when the effect size is non-collapsible (for example when treatment effects are logHRs or log odds ratios)~\cite{greenland1996absence}. The first two assumptions of normality of the treatment effects and standard errors used as within-study variances are assumptions which are commonly made in meta-analysis and further discussion of such assumptions is beyond the scope of this paper. The final assumption of linearity of the treatment effects however, is an assumption introduced to allow interpolation of treatment effects from biomarker-mixed populations. While this assumption is shown to be too strong in scenarios where the true systematic difference between biomarker-positive and biomarker-negative subgroups is large, in scenarios where the systematic difference is not very large, this assumption appears to be reasonable. 

\section{Conclusions}\label{section:conclusions}

We have described methods for incorporating treatment effects from studies with varying reporting of biomarker subgroup analyses in a single meta-analysis to estimate pooled treatment effects for a biomarker subgroup of interest. These methods have been applied to an illustrative example in mCRC and evaluated using a simulation study. We conclude that our method described to incorporate treatment effects from biomarker-positive, biomarker-negative and biomarker-mixed populations provides a consistent reduction in uncertainty around the estimated pooled treatment effect for biomarker-positive patients compared to using biomarker-positive data alone or biomarker-positive and biomarker-negative data. When the method is applied to treatment effects from biomarker-mixed populations which are not adjusted for biomarker status, the model is not robust to increasing values of the mean and variance of the systematic difference resulting in increases in bias. However, this effect is mitigated when applying the method to treatment effects from biomarker-mixed populations which are adjusted for biomarker status. We hope these methods and simulation study are informative for researchers seeking to conduct a pairwise meta-analysis of trials conducted in varying populations. 

\section*{Acknowledgements}
This research used the ALICE/SPECTRE High Performance Computing Facility at the University of Leicester. The authors thank Heather Poad for sharing data from the literature review on mCRC. 
This research was funded by the Medical Research Council, methodology research grant [MR/T025166/1]. Lorna Wheaton was also funded by National Institute for Health and Care Research (NIHR) Pre-doctoral Fellowship [NIHR301013] and by the Wellcome Trust Doctoral Training Programme [218505/Z/19/Z]. SB and LW were also supported by the NIHR Applied Research Collaboration East Midlands (ARC EM) and Leicester NIHR Biomedical Research Centre (BRC). The views expressed are those of the author(s) and not necessarily those of the NIHR or the Department
of Health and Social Care.

\bibliographystyle{unsrt}
\bibliography{main}

\begin{appendices}

\newpage

\setcounter{table}{0}
\renewcommand{\thetable}{A\arabic{table}}

\section{Prior Distributions for Proportions}

\textbf{Van Cutsem 2009}

\begin{table}[H]
\centering
\caption{Number of patients in control and treatment arms in Van Cutsem 2009 for WT, MT and Mixed analysis}
\begin{tabular}{@{}llll@{}}
\toprule
        & No. Control Patients & No. Treatment Patients & Total \\ \midrule
WT      & 350                  & 316                    & 666   \\
MT      & 183                  & 214                    & 397   \\
WT + MT & 533                  & 530                    & 1063  \\
Mixed   & 599                  & 599                    & 1198  \\ \bottomrule
\end{tabular}
\end{table}

\noindent
This paper reports KRAS status for 89\% of the biomarker-mixed population. Where KRAS status is reported the proportion MT patients = $397/1063 = 0.373$. 

\noindent
Using variance formula: 

\begin{equation}
    Var(p)=\frac{1(1-p)}{n}
\end{equation}

\noindent
Assuming $p=0.373$ and $n=1063$: 

\begin{equation}
    Var(p) = \frac{0.373(1-0.373)}{1063} = 0.00022
\end{equation}

\noindent
Assuming a mean of $0.373$ and a variance of $0.00022$ a beta distribution can be constructed using method of moments calculation where: 

\begin{equation}
    mean = m = \frac{\alpha}{\alpha + \beta}
\end{equation}

\begin{equation}
    var = \frac{m(1-m)}{\alpha + \beta + 1} 
\end{equation}

\noindent
After method of moments calculation, this gives a beta prior distribution of: 

\begin{equation}
    p \sim Beta(396, 666)
\end{equation}

\noindent
\textbf{Guren 2017}

\begin{table}[H]
\centering
\caption{Number of patients in control and treatment arms in Guren 2017 for WT, MT and Mixed analysis }
\begin{tabular}{@{}llll@{}}
\toprule
        & No. Control Patients & No. Treatment Patients & Total \\ \midrule
WT      & 97                   & 97                     & 194   \\
MT      & 58                   & 72                     & 130   \\
WT + MT & 155                  & 169                    & 324   \\
Mixed   & 185                  & 194                    & 379   \\ \bottomrule
\end{tabular}
\end{table}

\noindent
This paper reports KRAS status for 86\% of the biomarker-mixed population. Where KRAS status is reported the proportion MT patients = $0.401$. 

\begin{equation}
    p = \frac{130}{324} = 0.401 
\end{equation}

\begin{equation}
    Var(p) = \frac{0.401(1-0.401)}{324} = 0.000741
\end{equation}

\noindent
After method of moments calculation, this gives a beta prior distribution of: 

\begin{equation}
    p \sim Beta(129.6, 193.6) 
\end{equation}

\noindent
\textbf{Bokemeyer 2009}

\begin{table}[H]
\centering
\caption{Number of patients in control and treatment arms in Bokemeyer 2009 for WT, MT and Mixed analysis }
\begin{tabular}{@{}llll@{}}
\toprule
        & No. Control Patients & No. Treatment Patients & Total \\ \midrule
WT      & 97                   & 82                     & 179   \\
MT      & 59                   & 77                     & 136   \\
WT + MT & 156                  & 159                    & 315   \\
Mixed   & 168                  & 169                    & 337   \\ \bottomrule
\end{tabular}
\end{table}

\noindent
This paper reports KRAS status for 94\% of biomarker-mixed population. Where KRAS status is reported the proportion of MT patients = $0.431$. 

\begin{equation}
    p = \frac{136}{315} = 0.431
\end{equation}

\begin{equation}
    Var(p) = \frac{0.431(1-0.431)}{315} = 0.000779
\end{equation}

\noindent
After method of moments calculation, this gives a beta prior distribution of: 

\begin{equation}
    p \sim Beta(135.25, 178.56)
\end{equation}

\noindent
\textbf{Sobrero 2008 and Modest 2019}

These papers do not report the proportion of KRAS MT patients for any percentage of the population. Therefore, we define an informative beta prior distribution based on the prevalence of KRAS mutations in colorectal cancer. 

In a study of 1018 cases of metastatic colorectal cancer, Neumann et al ~\cite{neumann2009frequency} found KRAS mutations in 39.3\% of patients which supported previous research reporting KRAS mutations in 30-54\% of metastatic colorectal tumours. 

We assume the prevalence of MT patients is normally distributed such that: 

\begin{equation}
    Range = mean \pm 2 \times SD
\end{equation}

Therefore, we assume the mean prevalence is 0.42 and the standard deviation is 0.06. We then used method of moments to obtain the following beta distribution: 

\begin{equation}
    p \sim Beta(28, 38.67) 
\end{equation}

This beta prior distribution is for mixed studies when the proportion of MT KRAS patients is unknown. 

\section{Data for Sensitivity Analysis}

\setcounter{table}{0}
\renewcommand{\thetable}{B\arabic{table}}

\begin{table}[H]
\centering
\caption{LogHRs ($Y$) and corresponding standard errors ($\sigma$) on overall survival in KRAS WT ($+$), KRAS MT ($-$) and KRAS WT and MT ($mix$) populations from illustrative example in metastatic colorectal cancer used in sensitivity analysis. NA indicates no such treatment effect estimate is used from this study. }
\begin{tabular}{@{}lllllll@{}}
\toprule
Study           & $Y\_+$  & $\sigma_+$ & $Y_-$  & $\sigma_-$ & $Y_{mix}$ & $\sigma_{mix}$ \\ \midrule
Bokemeyer 2009  & -0.16 & 0.18                     & 0.25  & 0.20                     & NA     & NA                         \\
Ciardiello 2016 & -0.15 & 0.17                     & NA    & NA                       & NA     & NA                         \\
Douillard 2014  & -0.13 & 0.10                     & 0.16  & 0.11                     & NA     & NA                         \\
Guren 2017      & 0.13  & 0.18                     & 0.03  & 0.21                     & NA     & NA                         \\
Modest 2019     & NA    & NA                       & NA    & NA                       & -0.40  & 0.25                       \\
Peeters 2010    & -0.16 & 0.10                     & -0.06 & 0.11                     & NA     & NA                         \\
Peeters 2014    & -0.08 & 0.09                     & -0.07 & 0.10                     & NA     & NA                         \\
Primrose 2014   & 0.37  & 0.18                     & NA    & NA                       & NA     & NA                         \\
Qin 2018        & -0.27 & 0.12                     & NA    & NA                       & NA     & NA                         \\
Seymour 2013    & 0.01  & 0.10                     & NA    & NA                       & NA     & NA                         \\
Sobrero 2008    & NA    & NA                       & NA    & NA                       & -0.03  & 0.07                       \\
Van Cutsem 2009 & -0.18 & 0.14                     & 0.03  & 0.17                     & NA     & NA                         \\
Ye 2013         & 0.62  & 0.25                     & NA    & NA                       & NA     & NA                         \\ \bottomrule
\end{tabular}
\end{table}

\section{Model including studies reporting treatment effects from biomarker-negative patients only}\label{appendix:incl_bio_neg}

\setcounter{table}{0}
\renewcommand{\thetable}{C\arabic{table}}

While the illustrative example in mCRC did not contain any studies which reported treatment effects from biomarker-negative patients only, it is possible to include treatment effects from studies only investigating biomarker-negative patients in addition to studies investigating biomarker-positive patients only and studies investigating both biomarker-positive and biomarker-negative patients. 

The model for treatment effects from analysis of biomarker-positive patients only remains the same as described in equations (\ref{eq:withinWT}) and (\ref{eq:betweenWT}) in Section 3.1 of the main manuscript and the model for treatment effects from analysis of biomarker-positive and biomarker-negative patients remains the same as described in equations (\ref{eqn:within_MT})-(\ref{eqn:between_beta1}) in Section 3.2. However, we now define $n_-$ as the number of studies reporting treatment effects in biomarker-negative patients only. For studies $i=n_{+}+n_{\pm}+1,...,n_{+}+n_{\pm}+n_{-}$, the model is as follows:  

\begin{equation}
    Y_{-i}|\delta_{-i} \sim N(\delta_{-i}, \sigma^{2}_{-i})  
\end{equation}

\begin{equation} 
    \delta_{-i} = \delta_{+i} + \beta_{i}
\end{equation}

\begin{equation}
    \beta_{i} \sim N(\mu_{\beta}, \tau^{2}_{\beta}) 
\end{equation}

The prior distributions are the same as those described in Sections \ref{section:REMA_positive_only} and \ref{section:REMA_+-} in the main manuscript. 

While the example in mCRC cannot provide a data example for this model, the structure of the data required for this model can be seen in Table \ref{table:dummydata}.

\begin{table}[H]
\centering
\caption{Dummy example data to illustrate the structure of data required for the model including studies reporting treatment effects from biomarker-positive patients only, biomarker-negative patients only and biomarker-positive and biomarker-negative patients.}
\label{table:dummydata}
\begin{tabular}{@{}llll@{}}
\toprule
$Y_{+}$ & $\sigma_{+}$ & $Y_{-}$ & $\sigma_{-}$ \\ \midrule
-0.15       & 0.17         & NA          & NA           \\
-0.62       & 0.25         & NA          & NA           \\
0.01        & 0.10         & NA          & NA           \\
-0.27       & 0.12         & NA          & NA           \\
0.37        & 0.18         & NA          & NA           \\
NA          & NA           & -0.05       & 0.10         \\
NA          & NA           & -0.08       & 0.11         \\
NA          & NA           & -0.10       & 0.15         \\
-0.16       & 0.10         & -0.06       & 0.11         \\
-0.08       & 0.09         & -0.07       & 0.10         \\
-0.13       & 0.10         & 0.16        & 0.11         \\ \bottomrule
\end{tabular}
\end{table}

\section{Simulation Study Group 5 Line Plot}

\setcounter{table}{0}
\renewcommand{\thetable}{D\arabic{table}}
\setcounter{figure}{0}
\renewcommand{\thefigure}{D\arabic{figure}}

\begin{figure}[H]
  \includegraphics[width=\linewidth]{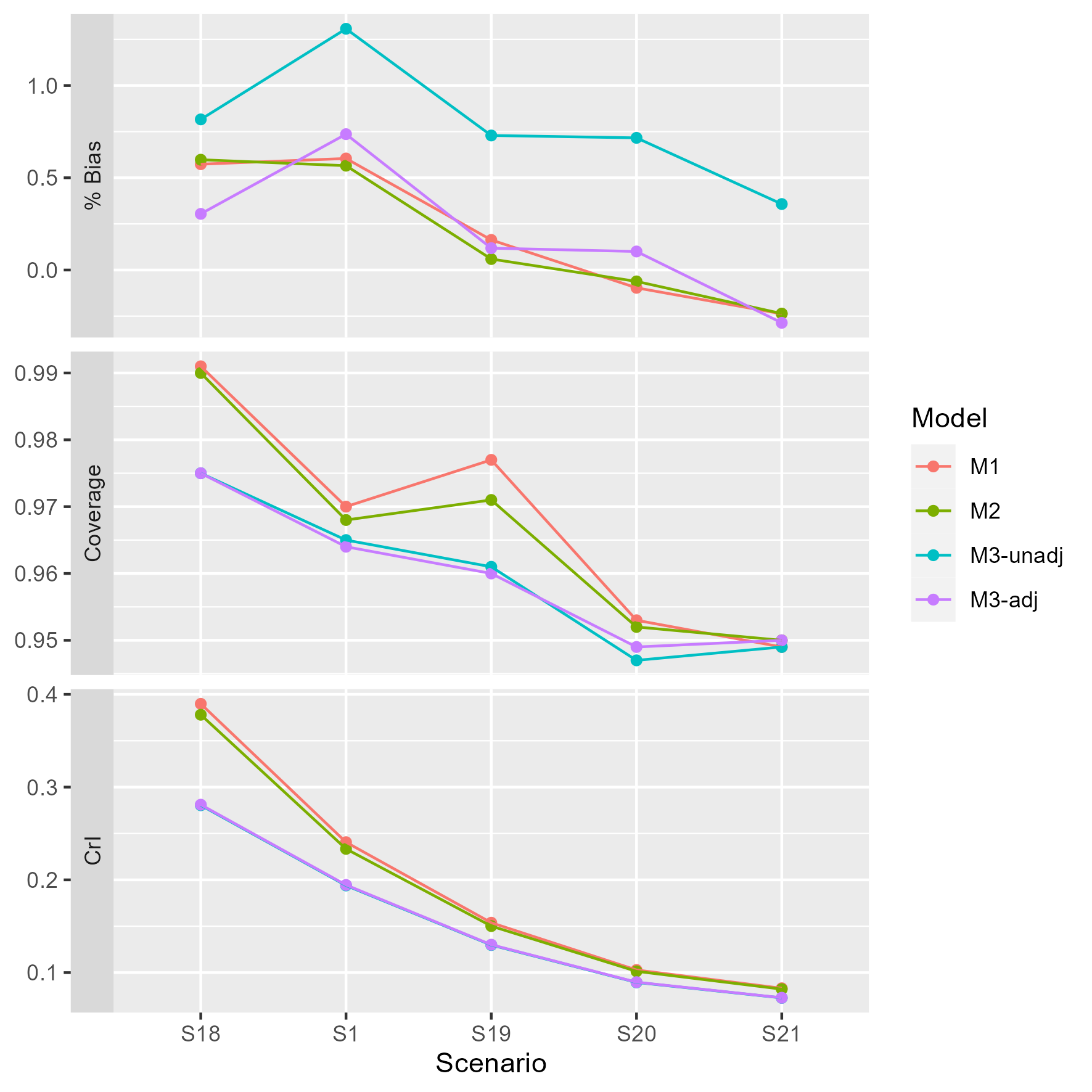}
  \caption{Percentage bias, coverage and mean credible interval width for pooled treatment effects for the biomarker-positive subgroup, across scenarios S18, S1 \& S21 where the total number of studies increases.} 
\end{figure}

\end{appendices}

\end{document}